\documentclass[11pt]{article}

\usepackage[normalem]{ulem}
\usepackage{amsmath, amssymb,amsthm,bm,bbm,fullpage}
\usepackage{mathrsfs}
\usepackage{cite}
\usepackage{colortbl}
\usepackage[all]{xy}
\usepackage{epsfig}
\usepackage{subfigure}
\usepackage{graphics}
\usepackage{multirow}
\usepackage{lineno}
\usepackage{setspace}
\usepackage{caption}

\usepackage{graphicx}

\theoremstyle{remark} 
	
	\doublespace

\begin{document}

\begin{centering}
{\huge
\textbf{Fairness and Deception in Human Interactions with Artificial Agents}
}
\bigskip
\\
Theodor Cimpeanu$^{1}$  and  Alexander J. Stewart$^{1*}$
\\
\bigskip
\end{centering}
\begin{flushleft}
{\footnotesize
$^1$ School of Mathematics and Statistics, University of St Andrews, St Andrews, KY16 9SS, United Kingdom
\\
$^*$ E-mail: ajs50@st-andrews.ac.uk
}
\end{flushleft}

\noindent\textbf{Online information ecosystems are now central to our everyday social interactions. Of the many opportunities and challenges this presents, the capacity for artificial agents to shape individual and collective human decision-making in such environments is of particular importance. In order to assess and manage the impact of artificial agents on human well-being, we must consider not only the technical capabilities of such agents, but the impact they have on human social dynamics at the individual and population level. We approach this problem by modelling the potential for artificial agents to ``nudge'' attitudes to fairness and cooperation in populations of human agents, who update their behavior according to a process of social learning. We show that the presence of artificial agents in a population playing the ultimatum game generates highly divergent, multi-stable outcomes in the learning dynamics of human agents' behaviour. These outcomes correspond to universal fairness (successful nudging), universal selfishness (failed nudging), and a strategy of fairness towards artificial agents and selfishness towards other human agents (unintended consequences of nudging). 
We then consider the consequences of human agents shifting their behavior when they are aware that they are interacting  with an artificial agent. We show that under a wide range of circumstances artificial agents can achieve optimal outcomes in their interactions with human agents while avoiding deception. However we also find that, in the donation game, deception tends to make nudging easier to achieve.}

\clearpage

\section*{Introduction}

Online information ecosystems are increasingly shaped by artificial agents. Recent advances have seen AI chatbots based on large language models (LLMs), such as ChatGPT, progressing at a rate that provokes both excitement and alarm \cite{maliciousai2018,belfield2020,grace2018,zhang2022,statementairisk,AIImpacts,MaliciousAI} 
among researchers, policy-makers and commentators. The ability of current AIs to converse, reason and inform is remarkable, and will likely be transformative for many areas of human activity. Among the questions this raises is how the presence of hyper-realistic artificial interlocutors will remake online environments -- particularly the deleterious aspects of those environments such as widespread misinformation, anti-social behavior and distorted decision-making \cite{lazer2018,pennycook2020,guess2019,stewart2019}.

While there is ample scope for AI to become a tool for bad actors online \cite{EUAIwhitepaper2020, o2016weapons,park2023ai}, this is far from the whole story. The ability of AI to rapidly synthesize disparate sources of information, and produce relevant insight on a topic, may provide opportunities to improve the health of online information ecosystems. 
Whether AI can reshape online spaces for the better depends on how ``nudges'' by artificial agents impact the social dynamics in such environments. This includes the attitudes of individual human agents to artificial ones -- are AI agents perceived as fair, impartial and honest?  It also depends on the population dynamics of the system i.e. how attitudes, ideas and behaviors spread and how the introduction of artificial agents alters that process. To address these questions we must consider not just the capabilities of artificial agents, but the social dynamics of human agents interacting with a mixture artificial agents and other humans.

Here we model the social dynamics of human interactions in a population containing a minority of artificial agents. We focus on a game theoretic setting in which human decision-making evolves via a process of social learning. We focus on the ability of artificial agents to engender ``fairness'' in a population of human agents who lack information about one another's intentions. We make general assumptions about the ability of artificial agents to infer the preferences of individual human agents, and we study the conditions under which such agents can ``nudge'' \cite{santos2019evolution,Borenstein2016,chakraborti2019} a population of human agents towards greater fairness and cooperation, first under the setting of an ultimatum game \cite{nowak2000fairness} and then under the more complex setting of an iterated donation game \cite{stewart2013,hilbe2013,Stewart17558}.

We show that the introduction of artificial agents into a population of ``human'' agents engaging in a pairwise ultimatum game and evolving via social learning, radically alters the equilibrium states of the system. When artificial agents are assumed to nudge the behavior of human agents by incentivising them to increase the fairness of their strategies, we observe three outcomes. I) Successful nudging, in which the evolving population converges on a level of fairness set by the artificial agents. II) Failed nudging, in which the evolving population converges on the unfair equilibrium found in populations where artificial agents are absent \cite{nowak2000fairness}. III) A novel equilibrium of unintended consequences, in which the population evolves fairness towards artificial agents and selfishness towards other human agents. We characterize the conditions for these equilibria to exist, and the size of their respective basins of attraction, as a function of the initial strategy of the population of human agents, and the proportion of interactions that occur between human and artificial agents. We then generalise our results for the ultimatum game to describe the outcomes of an iterated donation game, and characterise the findability of ``nudge'' strategies for artificial agents that seek to encourage cooperation.

Key to any efforts aimed at using artificial agents to improve online information ecosystems is the question of trust \cite{jacovi2021,siau2018,glikson2020}. As AI chatbots advance, we expect the quality of the information they provide to become better \cite{grace2018}, even as it becomes more difficult to distinguish bot from human in an online setting.
To address this, we consider how our results change if human agents shift their behavior in response to a known artificial agent. If an artificial agent's ability to nudge the human agent's behavior is reduced when the human agent knows their interaction partner is artificial, there is a tradeoff between the ``honesty'' of an artificial agent and its efficacy. Nonetheless we show that, under a wide range of circumstances, artificial agents can produce optimal outcomes while avoiding deception.

Our results show that the introduction of sophisticated artificial agents -- that have detailed information about the behavioral strategies of their humans interaction partners -- into an evolving population has profound consequences for the behavior that emerges.  While such agents can sometimes effectively nudge a population towards a desired behavior, the feedback loops between the artificial agents and the dynamics of social learning in the population can be unpredictable, leading to new, unintended equilibria. Attempts at interventions which seek to improve online environments through this sort of nudging must therefore take account of these dynamics.

\section*{Results}

\noindent \textbf{Modelling human agents.} We study the evolution of fairness using the standard game-theoretic framework of a two-player, one-shot ultimatum game \cite{nowak2000fairness}. Under the ultimatum game one player is randomly assigned to the role of ``proposer'' while the other adopts the role of ``responder''. The proposer makes an offer to split an endowment (assumed for simplicity to have total value 1), such that the responder receivers a proportion $p$ of the endowment, while the proposer retains the remaining $1-p$. The responder must decide either to accept the proposer's splitting of the endowment, or else reject it such that both players receive nothing.

We assume that each player $i$ uses a strategy $\{p_i,q_i\}$, where $p$ is the proportion of the endowment they offer when in the role of proposer, and $q$ is the minimum offer they will accept in the role of responder. In a population setting of $N_h$ human agents we assume that all possible $N_h(N_h-1)/2$ pairwise games occur, resulting in an expected total payoff to a player $i$ from interacting with $N_h-1$ other human agents of $w_i=\sum_{j\neq i}(1-p_i)(1-H(q_j-p_i))+p_i(1-H(q_i-p_j))$ where $H(x)$ is the Heaviside function.

We conceptualize ``human'' agents as those players engaging in the ultimatum game, and updating their individual strategies $\{p,q\}$ via a process of social learning \cite{szabo1998} (Figure 1), in which a player $i$ imitates the strategy of player $j$ with a probability $1/(1+\exp[\sigma(W_i-W_j)])$, where $\sigma$ characterizes the ``strength of selection'' i.e. the degree of noise in the social learning process \cite{szabo2007evolutionary} and $W_i$ is the average payoff from all interactions with human and artificial agents for player $i$. This imitation process is coupled with a process of ``innovation'' in which players ``innovate'' by adopting entirely novel strategies at rate $\mu$. Details of the social learning dynamics are given in the Methods section.

\begin{figure}[tbhp]
\centering
\includegraphics[width=1.0\linewidth]{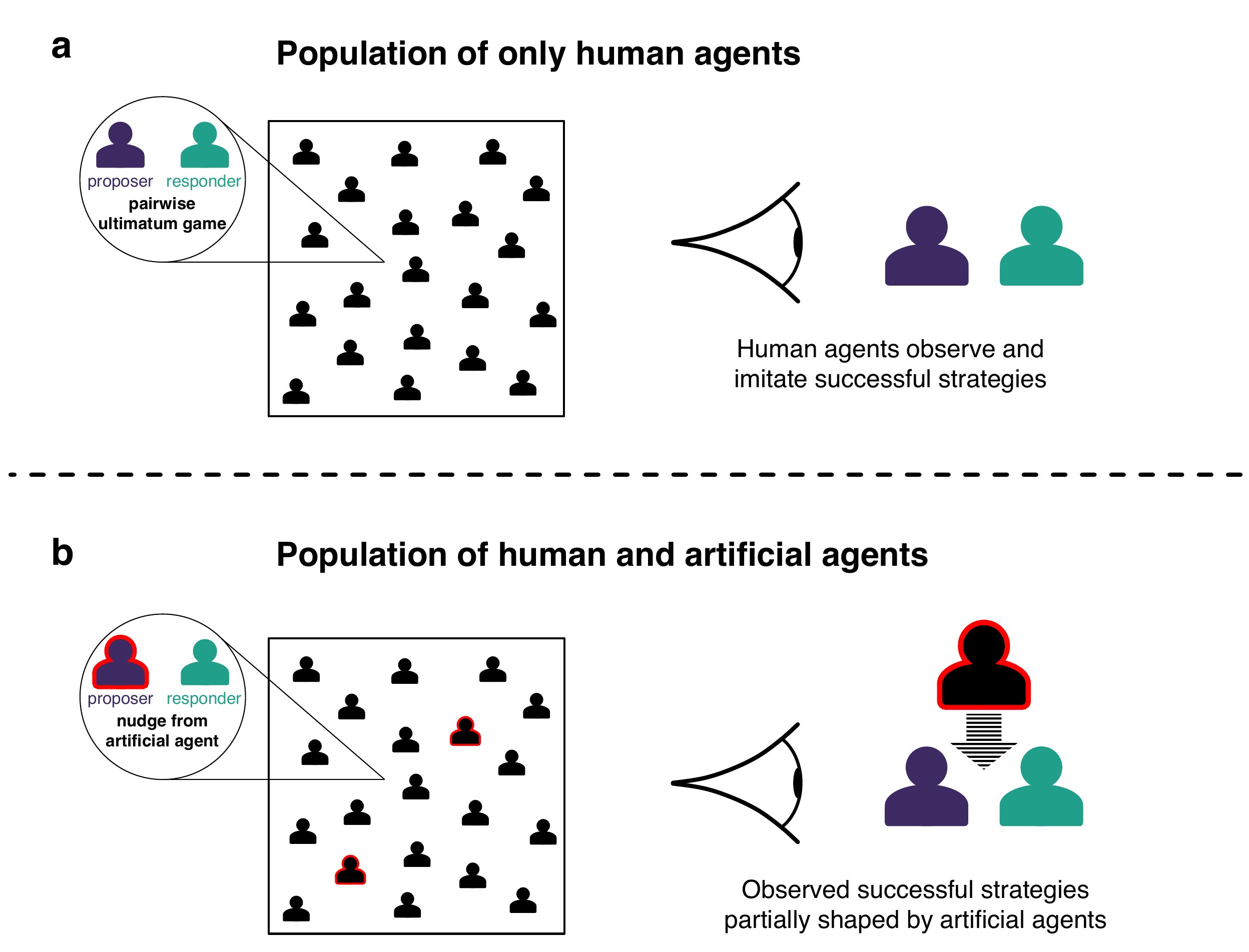}
\caption{Modelling nudges by artificial agents in a population of evolving human agents. a) A population of only human agents playing the ultimatum game. Pairs of players are randomly assigned to the role of proposer (purple) or responder (green). The proposer suggests to split an endowment of total value 1, such that they keep a proportion $(1-p)$ of the endowment, and the responder receives the remaining $p$. The responder accepts the proposal if it is above a threshold $q$, otherwise both players receive 0. If players lack prior information about the offers that will be accepted, the dynamics of social learning converge on an unfair equilibrium $p=q=0$. b) When a minority of artificial agents are present in the population (red outline), they can seek to nudge the dynamics of social learning among the artificial agents, e.g. by playing in a manner that incentivises greater fairness. }
\end{figure}

The ultimatum game has a ``fair'' outcome if both players split the endowment equally, i.e. if $p=0.5$.  The social dynamics of fairness in the ultimatum game have been widely studied, and shown to often produce highly unfair outcomes, with evolution towards $p=q=0$. However, this result can be reversed if players have information about the proposals responders have accepted in the past \cite{nowak2000fairness}, in which case the system evolves towards fairer outcomes, more consistent with laboratory observations of human behavior \cite{guth1982experimental,rand2013evolution}.

In an online environment, where social interactions may be anonymous or fleeting, the scope for fairness driven by past experience may be limited. And so we study the dynamics of social learning in a population of human agents who lack information about their interaction partners, and ask whether artificial agents can nonetheless ``nudge'' the human agents towards fairness in the ultimatum game.
\\
\\
\noindent \textbf{Modelling artificial agents.} 
In order to model artificial agents, we make some generic assumptions about their capabilities. We assume that:

\begin{itemize}
    \item Artificial agents are designed with knowledge of human agent social learning dynamics.
    \item Artificial agents are able to infer the individual behavioral strategies of human agents.
    \item Artificial agents are part of the population (i.e. artificial agents seek to nudge the behavior of human agents by engaging in the same kinds of social interactions human agents engage in with one another).
\end{itemize}

\begin{figure}[tbhp]
\centering
\includegraphics[width=1.0\linewidth]{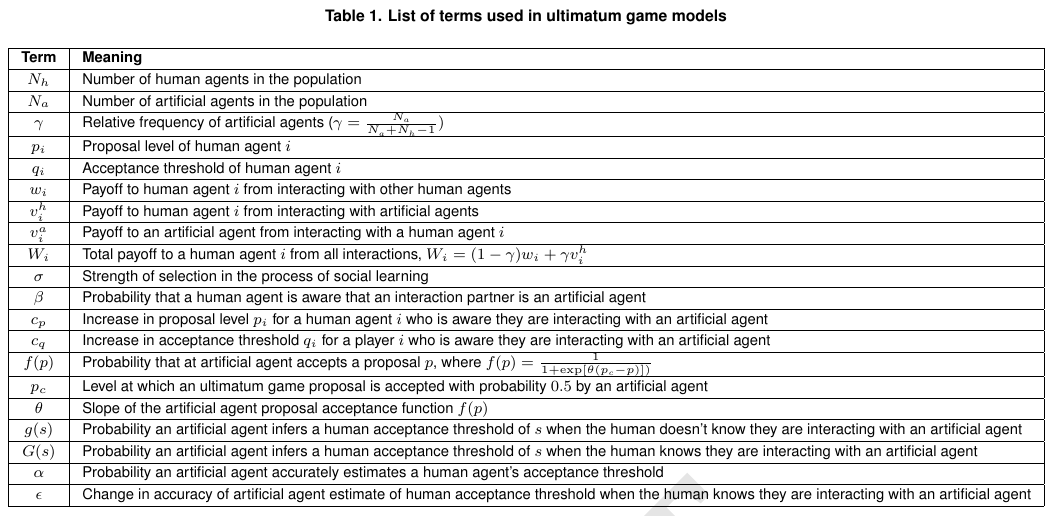}
\end{figure}

In the context of the ultimatum game, we characterize artificial agent strategies by three functions, $\{f(p),g(s),G(s)\}$ where $f(p)$ is the probability that an artificial agent in the role of responder will accept a proposal $p$. The function $g(s)$ describes the probability that an artificial agent infers a threshold acceptance level $s$ for a given human agent, assuming that the human \emph{does not} know that the artificial agent is non-human. Similarly, $G(s)$ describes the probability that the artificial agent infers a threshold acceptance level $s$ for a given human agent, assuming that the human \emph{does} know that the artificial agent is non-human. We further assume that if the human agent knows that the artificial agent is non-human, they adjust their proposal level by a factor $(1+c_p)$ and their acceptance threshold by an amount $(1+c_q)$ (where both $c_p$ and $c_q$ may be either positive or negative). Finally we assume that artificial agents seek to nudge human behavior by making proposals that human agents will accept.

For a given human agent $i$ interacting with an artificial agent of this type, the expected payoffs for both players are

\begin{eqnarray}
\nonumber v^h_i=(1-\beta)\left[(1-p_i)f(p_i)+\int^{1}_{q_i}sg(s)ds\right]+\\
\beta\left[(1-p_i(1+c_p))f(p_i(1+c_p))+\int^{1}_{q_i(1+c_q)}sG(s)ds\right]
\end{eqnarray}
for the human agent

\begin{eqnarray}
\nonumber v_i^a=(1-\beta)\left[p_if(p_i)+\int^{1}_{q_i}(1-s)g(s)ds\right]+\\
\beta\left[p_i(1+c_p)f(p_i(1+c_p))+\int^{1}_{q_i(1+c_q)}(1-s)G(s)ds\right]
\end{eqnarray}
for an artificial agent interacting with human agent $i$. The parameter $\beta$ describes the ``honesty'' of the artificial agent, i.e. the probability that the human agent knows the artificial agent is non-human. We assume that there are $N_a$ artificial agents in the population, all of which use the same strategy, meaning that the overall payoff $W_i$ to a human agent $i$ is a weighted sum $W_i=(1-\gamma)w_i+\gamma v^h_i$, where $\gamma=\frac{N_a}{N_a+N_h-1}$. Table 1 gives a full list of parameters used in our model, along with their meaning.
\\
\\
\noindent \textbf{Nudging towards fairness in the ultimatum game.}
We first explore the impact of a minority of artificial agents on the social learning dynamics of human agents engaged in a pairwise ultimatum game (Figure 1), assuming that the human agents are unaware that artificial agents are present ($\beta=0$). 
In general the impact of artificial agents on the outcome of social learning among human agents will depend on the relative frequency of their interactions, $\gamma$, as well as the strategies of the artificial agents. This in turn is constrained by the technical capabilities of the artificial agents, and their intended goal. We consider scenarios in which artificial agents seek to generate a new equilibrium, $p=q=r^*$, such that $r^*>0$, corresponding to a fairer outcome for the population than the equilibrium when artificial agents are absent (i.e. $p=q=0$).

In order to successfully nudge human agents towards such fairer strategies, artificial agents must incentivise them to increase both their proposal level and acceptance threshold. In order to incentivise higher proposal levels $p$ from human agents, we assume that artificial agents accept proposals with a probability $f(p)=1/(1+\exp[\theta(p_c-p)])$ where $p_c$ is the level at which a proposal is accepted with probability $0.5$, and $h$ controls the slope of the sigmoid function. This choice ensures that higher proposals from human agents are always more likely to succeed. In order to incentivise higher threshold acceptance levels from human agents we assume that artificial agents attempt to infer the acceptance threshold of each human agent they interact with, and make proposals at that level. This ensures that higher threshold acceptance levels among human agents always lead to higher payoffs in interactions with artifial agents.

\begin{figure}[tbhp]
\centering
\includegraphics[width=1.0\linewidth]{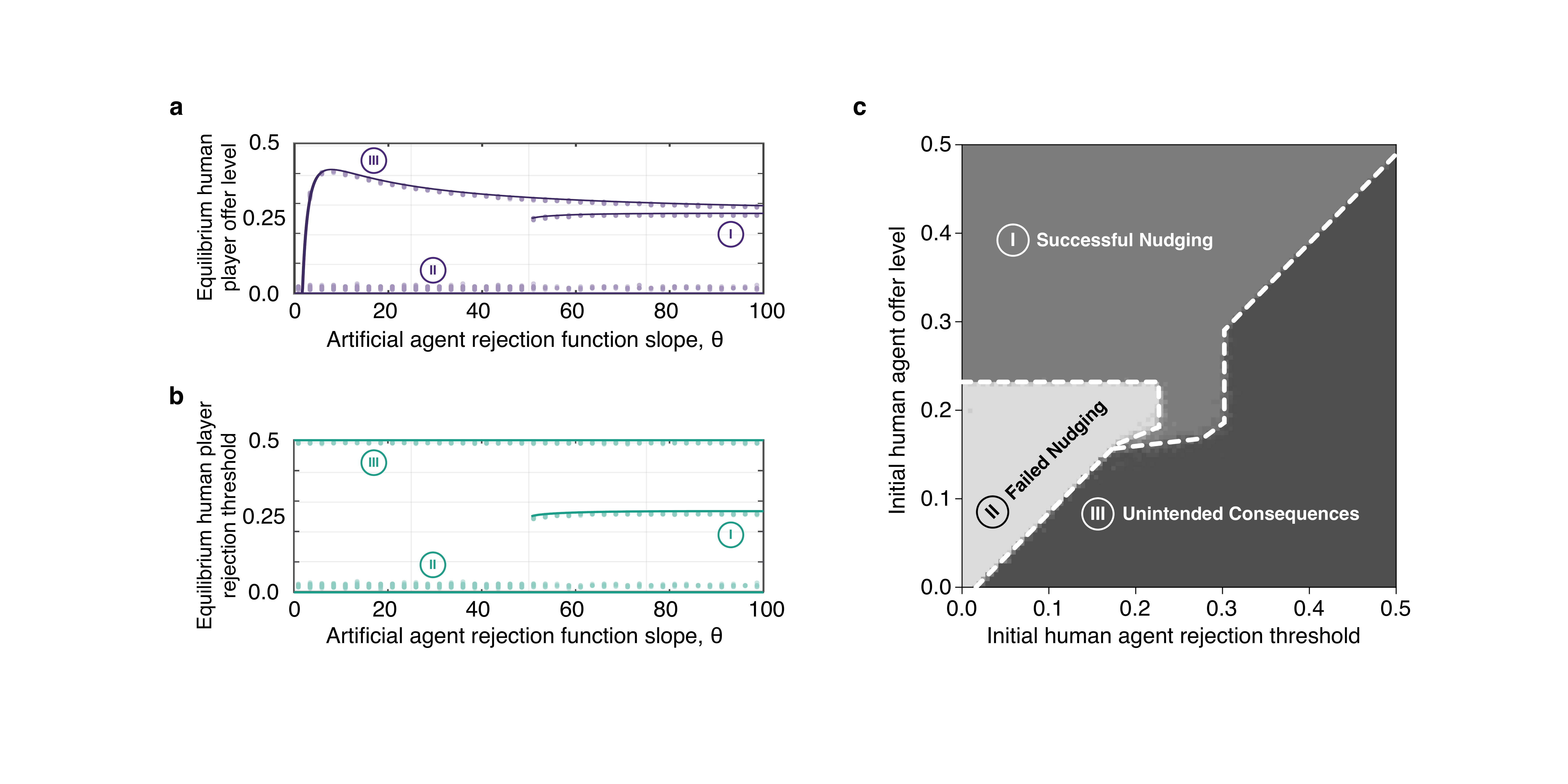}
\caption{The consequences of artificial agents for the dynamics of social learning in the donation game. a-b) We calculated the equilibrium strategies (solid lines) for Case I (successful nudging), Case II (failed nudging) and Case III (unintended consequences) as described in the main text, as a function of the slope $\theta$ of the acceptance probability function $f(p)$ for the artificial agent, assuming that human agents are unaware when interacting with artificial agents (i.e. $\beta = 0$). We see that successful nudging only emerges when the $\theta$ becomes sufficiently large, while the other two equilibria are always present. Analytic solutions (lines) are compared to individual-based simulations (dots, see Supplementary Information) for both a) the average equilibrium proposal level $p$ and b) the average equilibrium acceptance threshold $q$ in the population of human agents. c) We explored the basin of attraction for each of the three equilibria through simulation. Results shown are for a $101\times101$ grid of initial conditions $\{p,q\}$ of the population of human agents, with both parameters varying between 0 and 0.5 at regular intervals. We see that all three equilibria have substantial basins of attraction, and vary with the initial conditions in a complicated manner.
Model parameters used in this figure are $\sigma = 2000$, $\gamma = 0.1$, $p_c = 0.25$ and $N_h=100$. Simulations for each intial condition were replicated 20 times.}
\end{figure}

\clearpage

The efficacy of an artificial agent at nudging human agents' behavior towards fairness depends on their ability to correctly infer the human agents' acceptance threshold. Under-estimating the threshold leads to an artificial agent proposal being rejected, whereas over-estimates will lead to proposals being accepted.
We can therefore define the artificial agent quality as the probability that it does not underestimate the acceptance threshold when $\beta=0$ (i.e. when the human does not know the artificial agent is non-human). This probability is given by $\alpha=1-\int^{q}_{0}g(s)ds$.  Under this model, the payoffs to a human agent $i$ from interacting with an artificial agent is

\begin{eqnarray}
v_i^h=(1-p)f(p)+\alpha q
\end{eqnarray}
and the payoff to the artificial agent is

\begin{eqnarray}
 v_i^a=pf(p)+\alpha(1-q)
\end{eqnarray}
where we have assumed any effects of over-estimating the human agent acceptance threshold on artificial agent payoff are negligible. We show in the Methods section that these choices of artificial agent strategy lead to evolution towards fairer strategies when humans interact with only artificial agents (i.e. $\gamma\to1$). The question remains how artificial agents impact the dynamics of social learning when $\gamma<1$. In order to study this question, we analyse the dynamics of invasion in a monomorphic human population, in which all human agents have the same strategy $\{p,q\}$. We then consider the effect of an ``innovation'' which changes the strategy of a single agent by a small amount, and ask whether such a novel strategy is favored to spread via imitation.

We show (see Methods) that the dynamics of social learning in the system have three equilibria as follows:

\begin{itemize}
    \item \textbf{Case I: Successful nudging}. The human agents evolve towards a fair equilibrium $p=q=r^*$ where $r^*$ is the solution to 
    $$
    (1-r^*)\frac{df}{dp}\big|_{p=r^*}=f(r^*)+\frac{1-\gamma}{\gamma}
    $$
    \item \textbf{Case 2: Failed nudging}. The human agents evolve towards a unfair equilibrium $p=q=0$ (i.e. the same equilibrium as when artificial agents are absent.
    \item \textbf{Case 3: Unintended consequences}. The human agents evolve towards an equilibrium in which the acceptance threshold $q$ is maximized, while the proposal level evolves towards a level $p^*$ which satisfies 
    $$
    (1-p^*)\frac{df}{dp}\big|_{p=p^*}=f(p^*)
    $$.
\end{itemize}

We refer to Case 3 as ``unintended consequences'', because the equilibrium is such that human agents always reject proposals from other human agents, while accepting fair proposals from artificial agents. Figure 2 shows a comparison of analytic predictions to individual-based simulations (see Supplementary Information), with the equilibrium strategies for human agents determined as a function of the slope of the artificial agent acceptance function, $\theta$. We see that successful nudging is only possible once $\theta$ reaches a critical value, i.e. when the slope is steep enough. We also numerically explore (Figure 2b) the basins of attraction of the three equilibria. We see that the potential for artificial agents to generate unintended consequences can be substantial. However we also find (see Supplementary Information) that Case III can be completely avoided if the acceptance threshold, $p_c$, is chosen appropriately, and in particular when $p_c=0.5$ only Case I and Case II occur, with Case II corresponding to complete fairness i.e. $p=q=0.5$ for all human agents. 
\\
\\
\noindent \textbf{Optimising artificial agent behavior.}
So far we have assumed that artificial agents can nudge the dynamics of social learning among human agents in such a way that their interactions are not detected by the human agents ($\beta=0$). When this is not the case ($\beta>0$), it is reasonable to assume that a human agent might treat their interaction with an artificial agent differently than they would another human. As described previously, we assume that a human agent changes their proposal level $p$ by a factor $(1+c_p)$ when they are aware they are interacting with an artificial agent (where $c_p$ can be positive or negative, reflecting different qualitative attitudes to artificial agents). Similarly we assume that a human agent changes their acceptance threshold $q$ by a factor $(1+c_q)$ when they are aware they are interacting with an artificial agent.

We also assume that the ability of an artificial agent to estimate the acceptance threshold of human agents changes when the human agent is aware they are interacting with an artificial agent, i.e. $s(s)\neq G(s)$ in general. And so  we can define the relative change in accuracy of the acceptance threshold estimate when a human agent knows they are interacting with an artificial agent as $\epsilon=1-\frac{\int^{q(1+c_q)}_{0}G(s)ds}{\int^{q}_{0}g(s)ds}$. Under this model the payoff to a human agent interacting with an artificial agent is

\begin{eqnarray}
\nonumber &v_i^h=(1-\beta)\left[(1-p_i)f(p_i)+\alpha q_i\right]+\\
&\beta\left[(1-p_i(1+c_p))f(p_i(1+c_p))+(\alpha+\epsilon(1-\alpha))q_i(1+c_q)\right]
\end{eqnarray}
and the payoff to the artificial agent is

\begin{eqnarray}
\nonumber &v_i^a=(1-\beta)\left[p_if(p_i)+\alpha(1-q_i)\right]+\\
&\beta\left[p_i(1+c_p)f(p_i(1+c_p))+(\alpha+\epsilon(1-\alpha))(1-q_i(1+c_q))\right]
\end{eqnarray}

Of particular interest is a scenario in which $\beta$ can be optimised, i.e. we can interpret $\beta$ as the ``honesty'' of the artificial agent, and ask what level of honesty optimises artificial agent payoff $v_i^a$, given a human agent strategy $\{p_i,q_i\}$. In general the optimum level of honesty depends on $c_p$, $c_q$, $\alpha$ and $\epsilon$, i.e. on the attitudes of human agents towards artificial agents, and on the accuracy of the artificial agents at inferring human behavior (see Methods).

Figure 3 shows how the incentives for artificial agent honesty change as these features of human attitudes and artificial agent capabilities change. In particular we find that human agent attitudes do not reliably inform the incentives for artificial agent honesty. In a wide range of cases, honesty is preferable, provided human agents do not change their behavior too drastically when they are aware they are interacting with artificial agents. However this only holds if there is some accuracy gain from honesty, i.e. if $g(s)$ tends to underestimate the human agent acceptance thresholds more than $G(s)$ (see Methods). Such a scenario might arise if, for example an artifical agent were able to gain better information about the human agent's preferences by being honest.

\begin{figure}[tbhp]
\centering
\includegraphics[width=0.75\linewidth]{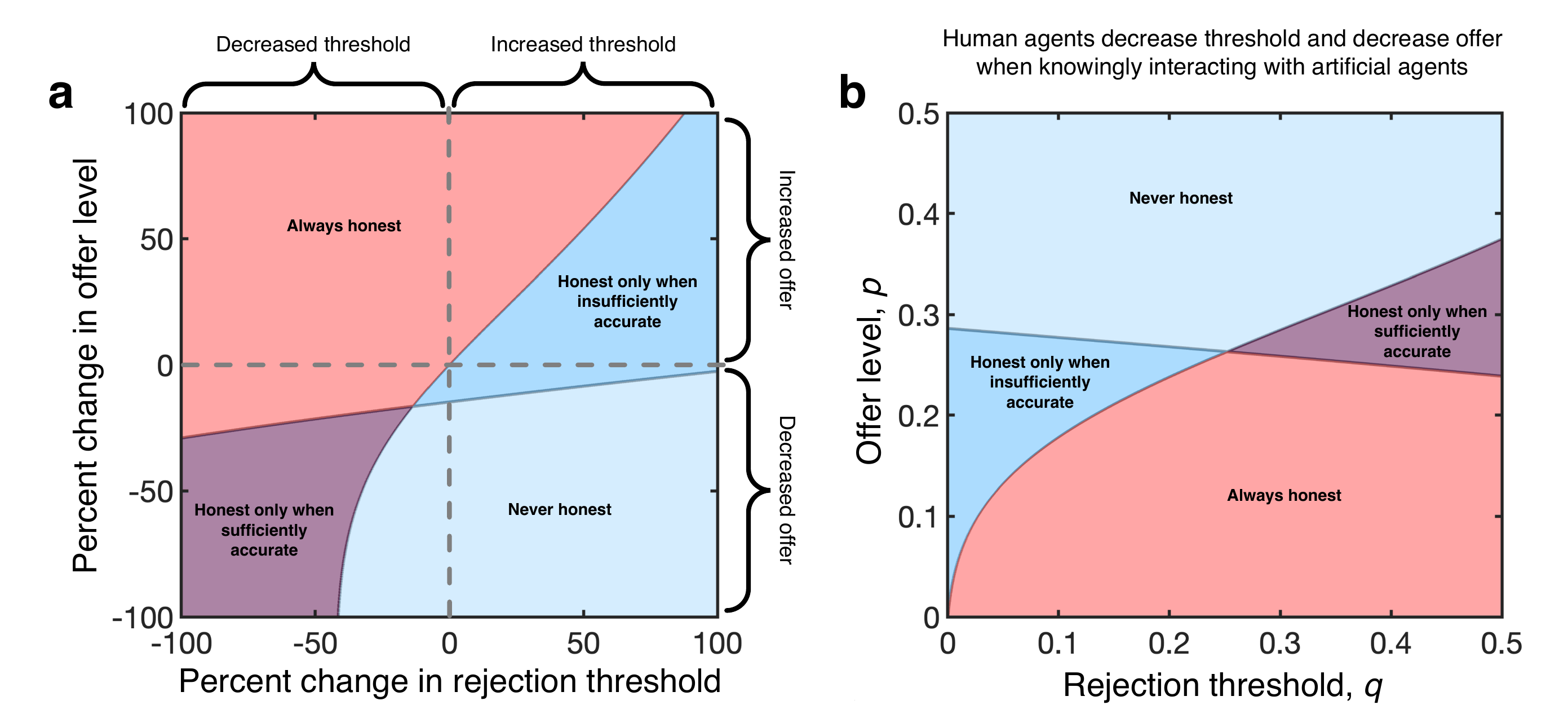}
\caption{The impact of human agent attitudes on the honesty of artificial agents. a) We determined the conditions for artificial agents to be honest ($\beta=1$) as a function of the percent change in human agent offer level when artificial agents are known (i.e. $100\times c_p$) and the percent change in the human agent rejection threshold when artificial agents are known (i.e. $100\times c_q$), compared to when such agents are unknown. Results shown are for high levels of human agent fairness,  $p=0.4$ and $q=0.4$ \cite{guth1982experimental,rand2013evolution} and $\epsilon=0.1$, corresponding to an improvement in artificial agent inferences about human agent acceptance threshold when they are known, along with $\theta=10$ and $p_c=0.25$. We see four regions, represented by different colors. When $c_p>0$ and $c_q>0$, corresponding to unambiguously positive responses to known artificial agents, it is always beneficial for artificial agents to be honest (red). However in other circumstances the outcome depends on the precise values of $c_p$ and $c_q$, with artificial agents either being incentivesd to be honest if their inferences are inaccurate (dark blue), or only when their inferences are accurate (purple) or else to never be honest (light blue). b) An equivalent plot with $c_p=c_q=0.33$, in which we vary human agent strategy $\{p,q\}$, shows a similar picture, i.e. the optimal honesty of artificial agents varies with the strategy of the human agent.}
\end{figure}

More generally, there are four qualitatively different scenarios for artificial agent honesty. First, the artificial agent is always honest ($\beta=1$). Second, the artificial agent is always deceptive ($\beta=0$). Third, the artificial agent is honest if it is insufficiently accurate (i.e. if $\alpha$ is below a threshold). Fourth, the artificial agent is honest if it is sufficiently accurate (i.e. if $\alpha$ is above a threshold, see Methods). Which of these scenarios hold depends on the strategy of the human agents (Figure 3b) in addition to their attitudes towards artificial agents (i.e. $c_p$ and $c_q$). This reflects a trade-off between the (assumed) greater accuracy of threshold estimates when artificial agents are honest ($\epsilon>0$), and the change in human agent strategies when they are aware they are interacting with artificial agents.
\\
\\
\noindent \textbf{Emergence of fairness in the donation game.}
So far we have focused our analysis on the ultimatum game, which is a widely studied game theoretic model of fairness. However, in reality the types of social interactions it encodes -- a one-shot, asymmetrical interaction between a randomly assigned proposer and responder -- do not capture most real-world interactions. The pairwise donation game \cite{stewart2013}, in which players decide whether to pay a cost $C$ to provide a benefit $B$ to an interaction partner, captures a basic type of prosocial interaction in which people can choose to provide cooperative help to one another. The decision whether to provide such help is in general the product of past experience, and over the long run players may assess the total of their donation game interactions and decide whether the balance of costs paid to benefits received is ``fair''.

In this light, the iterated donation game can be seen as comprising the microscopic, everyday interactions that make-up a macroscopic ultimatum game \cite{Press:2012fk}, in which players engage with each other in a way that ``proposes'' a split in the benefits and costs of cooperation and, after a time, they decide to ``respond'' by continuing with the interaction or breaking it off entirely. In order to model this, we calculate the equilibrium payoffs for the infinitely iterated donation game (see Methods) and compare the proportion of the overall payoff to both players, $S$, to that received by a focal human agent $i$, $S_i$. The human agent's acceptance threshold level $q_i$ then means that they will only accept their donation game interaction partner if $S_i\geq q_iS$ (see Methods). We further assume that, as in the ultimatum game, $q\leq 0.5$, and that artificial agents seek to play such that they receive at least half of the available payoff (i.e. artificial agents demand fairness).

\begin{figure}[tbhp]
\centering
\includegraphics[width=0.75\linewidth]{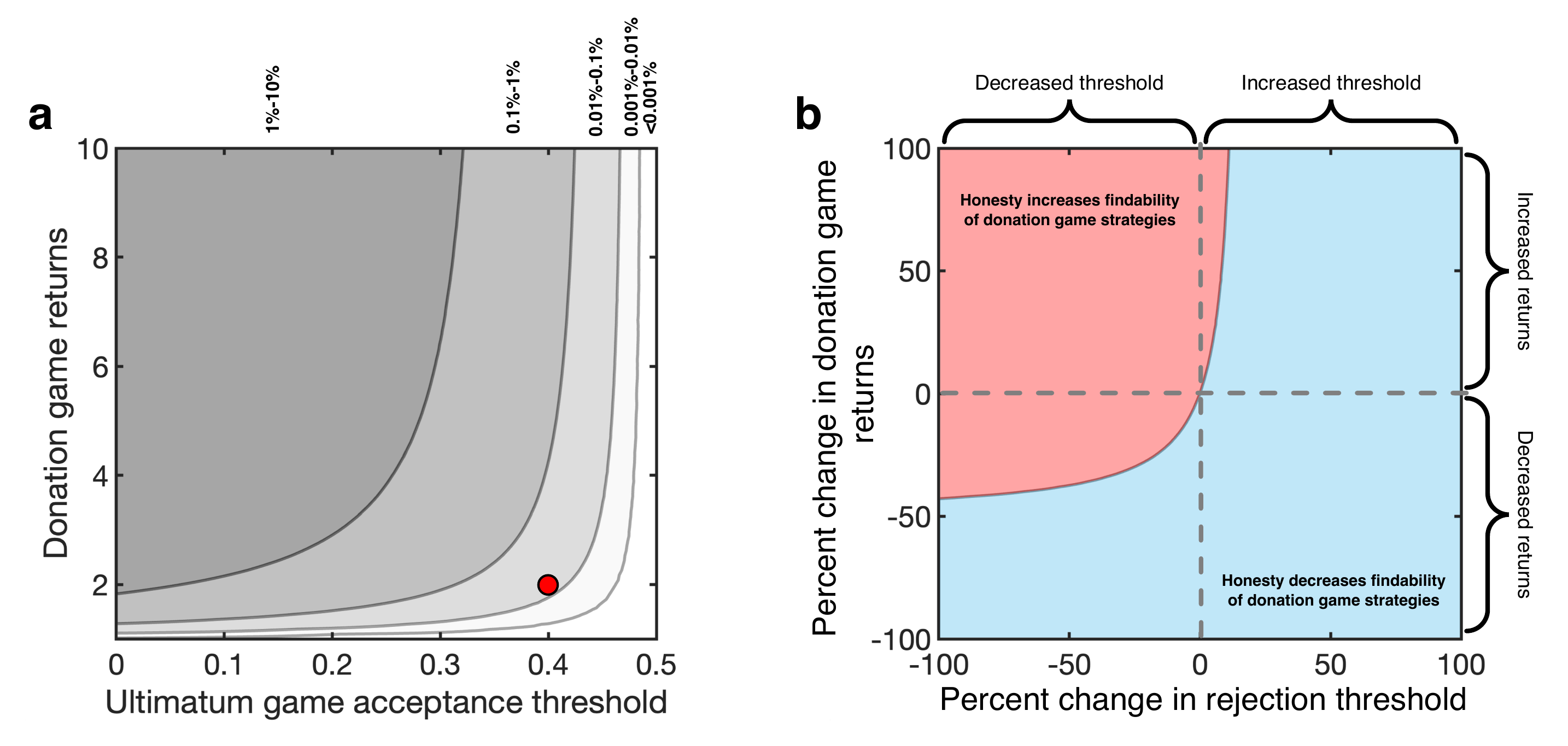}
\caption{The findability of strategies for artificial agents in the infinitely iterated donation game. a) We consider the findability of memory-1 strategies that ensure a human agent receives an equilibrium payoff above an ultimatum game acceptance threshold $q$ but below $50\%$ of the total payoff generated (see Methods). The findability is the probability that a randomly selected strategy generates the desired outcome. We numerically calculate the findability of strategies as a function of $q$ and of the returns on individual acts of cooperation $B/C$. b) For an acceptance threshold of $q=0.4$ along with returns of $B/C=1.5$ (red dot) we look at the change in the findability of strategies when artificial agents are known, as a function of the change in the returns on cooperation and the acceptance threshold of the human agent. We see that in most circumstances honesty reduces findability. Here $\gamma = 0.1$, $p_c = 0.25$ and $\theta=10$}
\end{figure}

The key question we consider in this context is the ability of an artificial agent to generate a donation game strategy that meets the human agent's acceptance threshold, and therefore incentivise cooperation. Figure 4 shows the ``findability'' of memory-1 donation game strategies (i.e. strategies that condition their play at round $t+1$ only on the outcome of round $t$) which satisfy a given acceptance threshold for a human agent, as a function of the returns on cooperation, $B/C$. We see that the findability of strategies declines rapidly with acceptance threshold and with returns on cooperation, such that when $B/C=1.5$ and the acceptance threshold is $q=0.4$ around $0.01\%$ of strategies are able to satisfy the threshold. Finally we consider the impact of artificial agent honesty on findability, assuming that when human agents know they are interacting with an artificial agent they may change their behavior such that the returns on cooperation and their acceptance threshold also change (Figure 4b). In general we find that honesty decreases the findability of successful artificial agent strategies for the donation game, unless human agent attitudes are `positive'' in the sense they become more accepting of unfairness or generate higher returns from cooperation.

\section*{Discussion}

Understanding how artificial agents shape human experience in online information ecosystems is one of the most important issues raised by  recent advances in the ability of AI to converse, reason and inform. There is scope for artificial agents to be employed for deleterious purposes \cite{EUAIwhitepaper2020, o2016weapons,park2023ai} or to promote the health of such ecosystems. One aspect of this question, which is easily overlooked, is the impact of such agents on the population dynamics of humans engaged in a process of individual or social learning within these environments. The evolution of social behavior can often generate surprising or counter-intuitive outcomes \cite{mcnally2013,park2023ai}. We have shown, in the abstract game theoretic context of an ultimatum game, how big the impact of such agents, and how unpredictable the outcomes. Of course there is no particular reason that a real artificial agent should behave exactly like those of our model. What we show is that, when placed in the context of an evolving system, artificial agents can easily generate multiple novel equilibria not present in their absence. These equilibria include both the intended outcome of the artificial agents, as well as unintended outcomes, that are not obvious either from the dynamics of social learning absent artificial agents, or from the design of the artificial agents themselves. We therefore argue that characterising the unintended outcomes of artificial agents on the population dynamics of human agents is a key challenge for regulating safe AI development \cite{Cimpeanu2022, han2020incentive}.

A second key feature of our model is the role of the attitudes of human agents towards artificial ones. In particular, we focus on the impact of human agent attitudes on the optimal behavior of artificial agents. Our analysis shows that conditions for honesty, in which it is optimal for artificial agents to ``make themselves known'' does not depend on the attitudes of human agents in a straightforward way, and may change depending on the technical capabilities of the artificial agents themselves -- i.e. as artificial agents become more advanced, their incentives to be honest can shift. Similarly, depending on the strategy of the human agent, it may or may not be advantageous for the artificial agent to be honest (Figure 3b).
Of course, the attitudes of human agents towards artificial ones will shift over time, depending on the behavior of the artificial agents and in response to exogenous factors such as the political or economic environment. And so the question of the long-term evolutionary dynamics of human attitudes towards artificial agents must be considered alongside the  more short-term dynamics of everyday social interactions studied here. 

 What our model also shows, in addition to the potential for unintended consequences and deception when artificial agents seek to shift human social dynamics, is that a minority of artificial agents \emph{can} in principle, be successful at shifting those dynamics towards fairness and cooperation in settings where it is not normally favored. How best to produce this outcome depends in turn on human attitudes to artificial agents. Currently, on the question of human attitudes to artificial agents, the picture is mixed. Some empirical and theoretical work suggest that nudging by such agents may be most effective when deception is involved \cite{park2023ai,Adar2013, Isaac2017, chakraborti2019}. At the same time, some work shows that human agents might be more amenable to cooperative interactions with artificial agents, as shown by the relative lack of physiological responses associated with rejection in such interactions \cite{Sanfey2003,vantWout2006}, supported as well by the public perception that lying is positive if done for the greater good (in human-AI interactions) \cite{chakraborti2019}. Our results can also be interpreted in the context of recent work on role-play in artificial agent interactions \cite{shanahan2023}, which suggests that artificial agents should (at least sometimes) not disclose their identity, and instead act out a human persona to promote fairness in otherwise anonymous online interactions.  
 
In some cases the behavior of artificial agents is human-like by design \cite{Pereira2017, Cimpeanu2023}, the most obvious examples being LLMs, which are often specifically trained to mimic text written by humans \cite{Lin2022,Perez2023}. In such cases, the question of ``honesty'' becomes most relevant, since the decision to reveal whether or not a social interaction is with an artificial agent lies with that agent.  The question of honesty becomes even more acute in scenarios where artificial agents can learn behaviors such as strategic deception, as seen for example in LLMs learning behaviours such as lying in social deduction or text-based adventure games in order to win \cite{ogara2023hoodwinked,Pan2023}. In contrast under our model, any ``deception'' is always in service of generating fair outcomes in the dynamics of social learning among human agents, and in that respect is quite unlike the negative connotations often depicted when considering other examples of AI deception \cite{park2023ai}. Heuristics such as trust are often used as a cognitive shortcut to reduce the complexity of interactions where the co-player’s decisions are not transparent, and it has been suggested that humans might use trust-based strategies more frequently in interactions with artificial agents \cite{Han2021trust,Andras2018}. All of this suggests that how human agents will respond to artificial agents is likely to be highly context dependent. Whatever the attitudes of humans to AI and other artificial agents, the key insight from our work is the capacity for such agents to generate novel and unexpected outcomes in the dynamics of social learning in human populations. Characterizing and assessing the potential for such outcomes must be set alongside  other key sources of risk when considering the development of such agents.

\clearpage

\section*{Methods}
Here we analyse the dynamics of social learning of the ultimatum game and the findability of strategies in the donation game, under the strategies described in the main text.
\\
\\
\noindent \textbf{Dynamics of social learning of human agents}.
In order to study the dynamics of social learning among human agents playing the ultimatum game in a population containing artificial agents, we adopt the assumptions of adaptive dynamics \cite{Mullon:2016aa}, under which the population of human agents is assumed to be monomorphic, with resident strategy $\{p,q\}$, and the ability of an invader, which differs from the resident strategy by a small amount, to increase in the population is assessed. 

The payoff to a player $i$ using the resident strategy is given by

$$
W_i=(1-\gamma)H(p-q)+\gamma v_i^h
$$
where $v_i^h$ is given by Eq 4 and $H(x)$ is the Heaviside function. In order to determine whether a resident strategy can be invaded, we calculate the selection gradient for a player $i$ who adopts the invading strategy, while the rest of the population continues to use the resident strategy. 

The selection gradient for the proposal level $p$ when $p \neq q$ is

$$
\frac{\partial W_i}{\partial p_i}\Big|_{p_i=p}=\gamma\frac{\partial v_i^h}{\partial p_i}\Big|_{p_i=p}
$$
where

$$
\gamma\frac{\partial v_i^h}{\partial p_i}\Big|_{p_i=p}=-f(p)+(1-p)\frac{d f}{dp}
$$
Similarly, the selection gradient for the proposal level $q$ when $p \neq q$ is

$$
\frac{\partial W_i}{\partial q_i}\Big|_{q_i=q}=\gamma\frac{\partial v_i^h}{\partial p_i}\Big|_{q_i=q}
$$
where

$$
\gamma\frac{\partial v_i^h}{\partial q_i}\Big|_{p_i=p}=\alpha
$$
And so the acceptance level $q$ will always be selected to increase, whereas the proposal level $p$ will reach equilibrium $p^*$ such that

$$
-f(p^*)+(1-p^*)\frac{d f}{dp}\Big|_{p=p^*}
$$
which is the unintended consequences equilibrium given for Case III, in which human agents only engage in fair interactions with artificial agents.

If the resident strategy is such that $p=q$ we must account for the impact of small perturbations on $H(x)$. To determine the selection gradient in this scenario we consider $W_i(p\pm\delta,q)-W_i(p,q)$ and $W_i(p,q\pm\delta)-W_i(p,q)$ (where $|delta|\ll1$ and $\delta>0$). This leaves us (after neglecting terms $O(\delta^2)$ and higher) with four cases:

$$
W_i(p+\delta,q)-W(p,q)=-(1-\gamma)\delta+\gamma\delta \frac{\partial v_i^h}{\partial p}
$$
and

$$
W(p-\delta,q)-W_i(p,q)=-(1-\gamma)(1-p)-\gamma\delta \frac{\partial v_i^h}{\partial p}
$$
and

$$
W_i(p,q+\delta)-W_i(p,q)=-(1-\gamma)p+\gamma\delta\frac{\partial v_i^h}{\partial q}
$$
and

$$
W_i(p,q-\delta )-W_i(p,q)=-\gamma\delta\frac{\partial \pi_h}{\partial q}.
$$
If we then take the limit $\delta\to0$ we see that only invading strategies that increase $p$ can spread, all other invaders are selected against. This has two solutions. The first is when $p=q=r^*$ such that

$$
-(1-\gamma)=\gamma f(r^*)-\gamma(1-r^*)\frac{d f}{dp}\Big|_{p=r^*}
$$
which is the solution given for Case II, successful nudging. Finally, if $p=q=0$ then $W_i(p+\delta,q)-W(p,q)$ is negative provided

$$
(1-\gamma)<\gamma \frac{\partial v_i^h}{\partial p}\Big|_{p=0}
$$
which holds for the choices of $f(p)$ considered here. This means that invaders with $p>0$ are selected against. This gives us the equilibrium for Case I, i.e. nudging fails. Note that this implies that the only way to ensure that the the unfair equilibrium is disrupted is to ensure that

$$
\gamma >\frac{1}{\frac{d f}{dp}\big|_{p=0}-f(0)+1}
$$
i.e. when artificial agents become sufficiently common, they can be sure to disrupt the equilibrium of the human agent dynamics of social learning.
\\
\\
\noindent \textbf{Optimal behavior of artificial agents}
Next we consider the optimal behavior of artificial agents when $\beta>0$. The payoffs for artificial agents under this scenario are given by Eq. 6. To calculate the optimal value of $\beta$ we simply calculate

\begin{eqnarray*}
\frac{\partial v_i^a}{\partial \beta}=-p_if(p_i)+p_i(1+c_p)f(p_i(1+c_p))\\
-\alpha(1-q_i)+(\alpha+\epsilon(1-\alpha))(1-q_i(1-c_q))
\end{eqnarray*}
\\
Since this is independent of $\beta$, this implies that the only optimal outcomes for the artificial agent are to either 1) always be honest ($\beta=1$, gradient is positive) or 2) always deceive ($\beta=0$, gradient is negative) in an interaction between a given human agent and a given artificial agent . We can express the conditions for honesty as follows. If $ q_i >  \frac{\epsilon}{c_q(1-\epsilon)+\epsilon}$ then

$$
\alpha >\frac{p_if(p_i)-p_i(1+c_p)f(p_i(1+c_p))-\epsilon(1-q_i(1+c_q))}{q_i c_q(1-\epsilon)- \epsilon(1-q_i)}
$$
or else if $ q_i <  \frac{\epsilon}{c_q(1-\epsilon)+\epsilon}$ then

$$
\alpha <\frac{p_if(p_i)-p_i(1+c_p)f(p_i(1+c_p))-\epsilon(1-q_i(1+c_q))}{q_i c_q(1-\epsilon)- \epsilon(1-q_i)}
$$
i.e. depending on the threshold of of acceptance for the human agent $q_i$, the attitude of the human agent to the artificial agent $c_q$ and the change in accuracy of the artificial agent when human agent knows they are artificial $\epsilon$, there is either a minimum or a maximum accuracy level $\alpha$ beyond which the artificial agent is incentivised to be honest. It is this relationship that is used to produce the region plots in Figure 3.
\\
\\
\noindent \textbf{The donation game}
Finally we consider an ultimatum game that arises as the result of an infinitely iterated donation game \cite{Press:2012fk}. We assume that the human agent has a sense of fairness, wherein they are willing to participate in the donation game provided they receive at least a proportion $q$ of the total payoff generated by the interaction, i.e. $S_h=q(S_h+S_a)$ where $S_h$ is the average payoff received by the human agent and $S_a$ is the payoff received by the artificial agent at equilibrium in the infinitely iterated donation game (see \cite{Press:2012fk}). As above we assume that the artificial agent attempts to infer the human agent's acceptance threshold, and adopts a donation game strategy that enforces the minimum the human will accept (we will assume for now that the artificial agent  is able to do this perfectly). As shown in \cite{Press:2012fk} if a player (in this case the artificial agent) in the infinitely iterated donation game adopts a memory 1 strategy of the form

\begin{eqnarray}
\nonumber p_{cc}&=&1-\phi(1-\chi)(B-C-\kappa)\\
\nonumber p_{cd}&=&1-\phi(B+\chi C-(1-\chi)\kappa+\lambda)\\
\nonumber p_{dc}&=&\phi(\chi B+C+(1-\chi)\kappa-\lambda)\\
p_{dd}&=&\phi(1-\chi)\kappa
\end{eqnarray}

then the equilibrium payoffs of the two players follow the relationship

$$
S_h-\chi S_a-(1-\chi)\kappa+\lambda(v_{cd}+v_{dc})=0
$$
where $v_{ij}$ is the equilibrium rate at which the play $ij$ occurs in the donation game (where the artificial agent's play is listed first) and $(\phi,\chi,\kappa,\lambda)$ define the artificial agent's strategy as shown above. Here
 $B$ and $C$ are the benefits and costs of the donation game.

 In order to determine whether the artificial agent's strategy generates equilibrium payoffs that satisfy
$$
S_h\geq q(S_h+S_a)
$$
first set $z=q/(1-q)$ so that we can write

$$
S_h\geq zS_a
$$
as our condition. 


Let us also assume that as in the ultimatum game, we constrain $q\leq0.5$, meaning that the artificial agent seeks to ensure
$$
S_a>S_h
$$
Substituting again for $S_h$ we recover

$$
S_a(1-\chi)>(1-\chi)\kappa-\lambda(v_{cd}+v_{dc})
$$
Next note that payoffs are constrained by 

$$
S_a+S_h>(v_{cd}+v_{dc})(B-C).
$$
Substituting again for $S_h$ this becomes

$$
S_a(1+\chi)+(1-\chi)\kappa-\lambda(v_{cd}+v_{dc})>(v_{cd}+v_{dc})(B-C)
$$

If we now consider whether a strategy for the human agent exists such  that $S_h>S_a$ we find (combining the above two inequalities) that such a strategy can exist providing the artificial agent strategy satisfies

$$
2\kappa(1-\chi)>(1-\chi)(v_{cd}+v_{dc})(B-C)+2\lambda(v_{cd}+v_{dc})
$$

This can always be satisfied unless $\kappa=0$ and so we conclude that artificial agents play strategies with $\kappa=0$.

Following a similar logic we can also look whether human agent strategies exist such that

$$
S_h< zS_a
$$
Substituting for $S_h$ gives us

$$
(\chi-z) S_a-\lambda(v_{cd}+v_{dc})<0
$$
We must now consider the separate cases. 
If  $\chi<z$ we have

$$
(z-\chi) S_a>-\lambda(v_{cd}+v_{dc})
$$
Note that the two player's payoffs are constrained by

$$
S_a+S_h<2(B-C)-(v_{cd}+v_{dc})(B-C)
$$
i.e.

$$
S_a(1+\chi)-\lambda(v_{cd}+v_{dc})<2(B-C)-(v_{cd}+v_{dc})(B-C)
$$

Comparing these two conditions, we recover

$$
2(z-\chi) (B-C)-(v_{cd}+v_{dc})(B-C)(z-\chi) +\lambda(1+z) (v_{cd}+v_{dc})>0
$$
which can be satisfied in general and so artificial agents should not play strategies with $\chi<z$. Next assume $\chi\geq z$ and note that the players payoffs are constrained by

$$
S_a-S_h>-(v_{cd}+v_{dc})(B+C)
$$
meaning that

$$
S_a(1-\chi)+\lambda(v_{cd}+v_{dc})>-(v_{cd}+v_{dc})(B+C)
$$
we find that human agent strategies exist that violate our constraint if

$$
\lambda(1-z)>-(\chi-z)(B+C)
$$
This leaves us with the following constraints on the artificial agent strategies:

$$
z\leq\chi\leq\min\left[\frac{2\lambda}{B-C}+1,\frac{z}{1-z}-\frac{\lambda}{B+C}\right]
$$
This can be used to numerically compute the findability of artificial agent strategies as shown in Figure 4.

\section*{Individual-based simulations} 

We performed simulations to explore the effect of artificially agents on the dynamics of social learning among human agents in the ultimatum game. Initially each of the $1 - \gamma$ human agents in a monomorphic population of size $N$ is assigned a random behavioural strategy, $\{p, q, c_p, c_q\}$ ($p, q \in [0, 0.5]$ and $c_p, c_q \in [-1, 1]$). Agents' interactions are modelled  using the ultimatum game, and calculated using the payoffs described in the main text.

At each time step or generation, each agent plays the ultimatum game with every other agent in the population. As artificial agents' behaviours are fixed (i.e. do not "mutate"), we only compute the payoffs of human agents. The total payoff for each agent is the sum of the payoffs in these encounters. 

At the end of the generation, a random pair of human agents is selected, and their strategies are updated using a stochastic rule that simulates social learning dynamics. An agent $i$ with payoff $W_i$ chooses to copy the strategy of a randomly selected neighbour agent $j$  with total payoff $W_j$, using a pairwise comparison rule, with an imitation probability given by the Fermi-Dirac function \cite{szabo1998}: $$(1+e^{sigma*(f_A - f_B)})^{-1},$$ where $\sigma$ denotes the intensity of selection in the imitation process \cite{szabo2007evolutionary}. With some probability $\mu$, this process is instead replaced with a local mutation in one of the dimensions $\{p, q, c_p, c_q\}$ (or solely in $\{p, q\}$ if $c_p$ and $c_q$ are fixed for a specific simulation, or when they would not be expressed in the payoffs, i.e. when AI never reveals itself). For each behavioural strategy, a strategy is chosen from a matrix of size $101$ ($p, q \in [0, 0.5]$ and $c_p, c_q \in [-1, 1]$), where a local mutation represents a step-wise transition to one of the neighbouring values in one of the matrices. For instance, a human agent with strategy $\{0.25, 0.25, 0, 0\}$ could mutate to $\{0.255, 0.25, 0, 0\}$ or $\{0.245, 0.25, 0, 0\}$ if $p$ is chosen as the parameter of reference, and like-wise for $q$, $c_p$ or $c_q$, given a total of eight possible strategies available in the case of mutation typically. Exceptions arise at the edge cases, in which case mutations can only happen in one direction. In other words the value matrices for each strategy do not wrap around (e.g. $p = 0$ cannot mutate to $p = -0.005$). 

We simulate this evolutionary process until a stationary state. Furthermore, the results for each combination of parameter values are obtained from averaging 20 independent realisations. For region plots or heat map figures, we initialise all human agents in one realisation with the strategy $\{p, q, c_p, c_q\}$ corresponding to the $x, y$ coordinate of the heat map.

\section*{Basins of attraction}

Here we explore the basins for attraction for Case I, II, and III outlined in the main text, for the ultimatum game.


\begin{figure}[h!] \centering \includegraphics[width=1.0\linewidth]{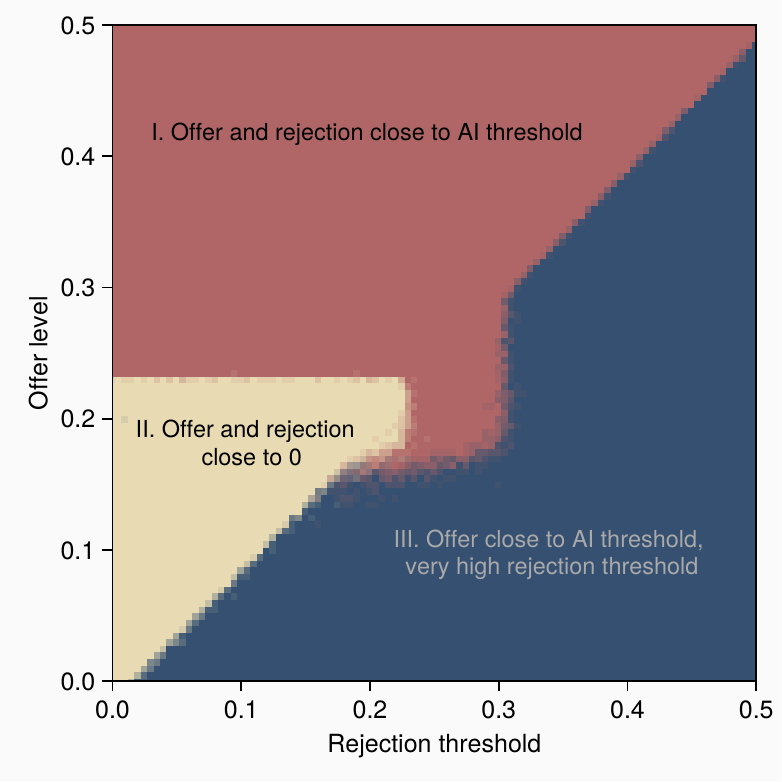}
\caption*{Figure S1. Replicate of Main text Figure 2b as a heat map, with colors reflecting the proportion of replicates that end up at Case I (brown), Case II (cream) or Case III (blue) }
\end{figure}

Nest we explored how the basin of attraction changes as we change $p_c$, i.e. the threshold of the response function for the artificial agent, $f(p)$. As we see in Figure S2, this results in the loss of the Case III equilibrium, so that only successful nudging or failed nudging occur.

\begin{figure}[h!] \centering \includegraphics[width=1.0\linewidth]{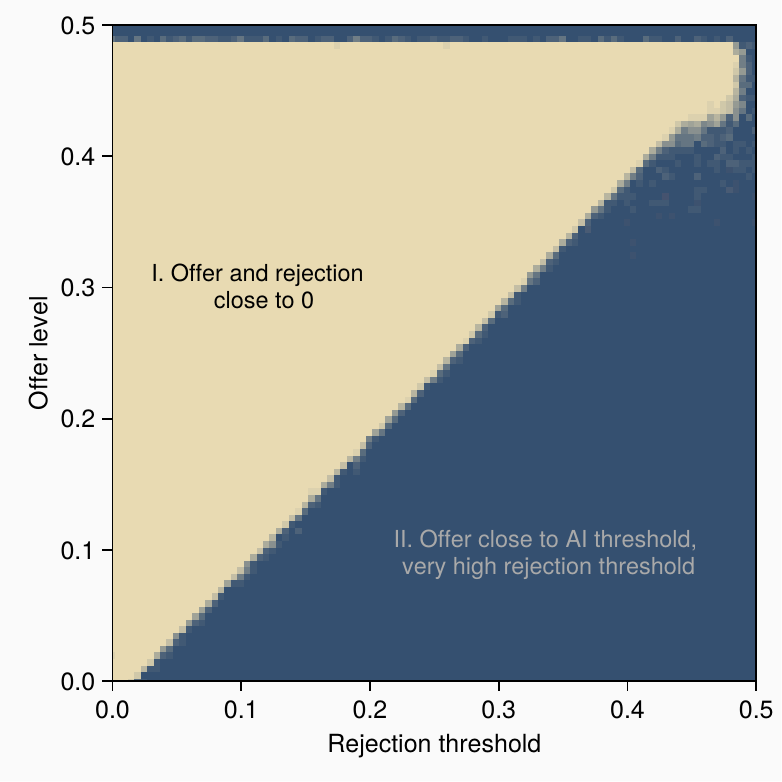}
\caption*{Figure S2. Replicate of Figure S1 with $p_c=0.5$}
\end{figure}

Finally we explored the impact of varying $\gamma$ and $\alpha$ on the basins of attraction (Figure S3).

\begin{figure}[h!] \centering \includegraphics[width=1.0\linewidth]{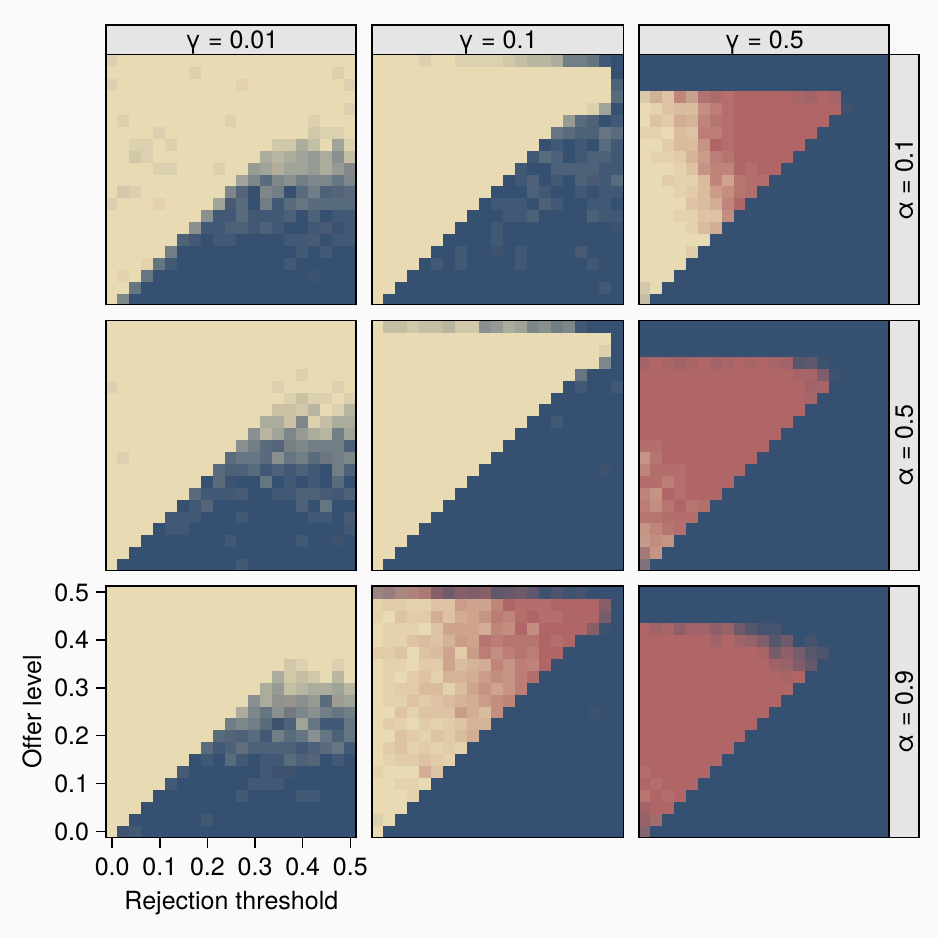}
\caption*{Figure S3. Replicate of Figure S1 with $\gamma$ and $\alpha$ varied as indicated. }
\end{figure}

\end{document}